# Check Your Data Freedom:
# A Taxonomy to Assess Life Science Database Openness

Melanie Dulong de Rosnay
Fellow, Science Commons and Berkman Center for Internet & Society at Harvard University

**Abstract**

Molecular biology data are subject to terms of use that vary widely between databases and curating institutions. This research presents a taxonomy of contractual and technical restrictions applicable to databases in life science. It builds upon research led by Science Commons demonstrating why open data and the freedom to integrate facilitate innovation and how this openness can be achieved. The taxonomy describes technical and legal restrictions applicable to life science databases, and its metadata have been used to assess terms of use of databases hosted by Life Science Resource Name (LSRN) Schema. While a few public domain policies are standardized, most terms of use are not harmonized, difficult to understand and impose controls that prevent others from effectively reusing data. Identifying a small number of restrictions allows one to quickly appreciate which databases are open. A checklist for data openness is proposed in order to assist database curators who wish to make their data more open to make sure they do so.

**Keywords**

Life science data, open access, bioinformatics, copyright law, database, taxonomy, Science Commons, public domain

**Introduction**

This work is being developed under the auspices of the Science Commons Data project and builds upon the Science Commons Open Access Data Protocol proposing requirements for interoperability of scientific data. Legal simplicity and predictability can be achieved by waiving copyright and other contractual restrictions, allowing data integrators to reuse, modify and redistribute large datasets, towards the freedom to integrate[1]. Legal accessibility issues are not the only hurdle to data integration. Technical accessibility[2] should be ensured in order to allow scientists to download data easily and use them in any way, including ways that the initial creators had not considered.

The objective of this research is to assess the accessibility of databases by analysing their access interfaces and their reuse policies. After presenting the rationale for sharing data in life science, we will qualify databases openness by analyzing a set of technical access interfaces and legal terms of use. A taxonomy of technical and legal restrictions will be presented and used to assess a sample of databases. Based on these criteria, we will propose a set of questions for database curators to assess their own data technical and legal openness.

---

[1] Wilbanks John, "Public domain, copyright licenses and the freedom to integrate science", *Journal of Science Communication*, volume 07, issue 02, June 2008.

[2] Defined as "Technical Open Access" in Dulong de Rosnay Melanie, Opening Access in a Networked Science, *Publius Project, Essays and conversations about constitutional moments on the Net collected by the Berkman Center*, June 2008.

## 1. The context of data sharing

The Human Genome project developed standards for rapid data sharing, allowing the scientific community to access and reuse data in other ways than the team at the origin of the first sequencing of the data. According to NIH-DOE [Guidelines](Guidelines) for Access to Mapping and Sequencing Data and Material Resources, "a key issue for the HGP (Human Genome Project) is how to promote and encourage the rapid sharing of material and data that are produced, especially information that has not yet been published or may never be published in its entirety. Such sharing is essential for progress toward the goals of the program and to avoid unnecessary duplication. It is also desirable to make the fruits of genome research available to the scientific community as a whole as soon as possible to expedite research in other areas."

## 2. Methodology

This research started with analyzing terms of use of databases from the Molecular Biology Database hosted by the Nucleic Acids Research [Journal](Journal), and assessing them regarding open access criteria as described in the [Science Commons Open Data Protocol](Science Commons Open Data Protocol). A policies sample has been retrieved and analyzed. The next step has been to identify barriers to open access and reuse of data from these database policies, and to build a taxonomy of restrictions. These restrictions can be of legal or contractual nature, but they can also be technical, e.g. the impossibility of downloading the whole database if its results can be accessed only through a field-based search. A systematic analysis of more database policies hosted by the Life Science Resource Name ([LSRN](LSRN)) Schema registry allowed us to confirm this taxonomy and to refine it by adding other terms.

In this section, we will define technical and legal accessibility conditions and restriction values. The purpose of identifying controls and restrictions applicable to databases is twofold. First, it will allow us to understand terms of use and other requirements governing the access to molecular biology databases and especially which control prevent the free sharing of data Second, these restrictions will be clustered into classes, making it possible to systematize the analysis of databases and to easily identify the data that are in the public domain and therefore can be reused by the scientific community.

### 2.1 The design of a taxonomy

Two types of control can be exercised on databases: technical restrictions embedded in the design of the database, and legal restrictions expressed in the terms of use.

Technical restrictions affect databases that cannot be searched or processed in any possible way. Technical openness is ensured by the possibility of downloading the whole dataset and reusing and integrating data, in the same way the Science Commons Neurocommons project provides a datamining platform allowing machine-readable representation and interpretation of data, or that Basic Local Alignment Search Tool ([BLAST](BLAST)) finds similarities between sequences. Semantic web processing applied to scientific data should improve the way science is performed and allow network effects by connecting knowledge from various datasets. Databases that require registration before access, or offer only a batch processing or a query-based mechanism to retrieve data after a specific search, do not comply with the technical requirements necessary to make data open.

Terms of use, licenses or access policies are legal texts describing authorized and unauthorized usages. Those legal rules are expressed by the entity distributing a product such as software or scientific data. The infringement of those self-declared rights can lead to lawsuits. Terms of use can be difficult to understand, even for lawyers, while scientists need to know quickly whether they can use a dataset.

Therefore, a set of questions has been designed to understand whether databases are, in fact, in the public domain and whether the data can be reused, redistributed and integrated.

---

Technical accessibility

*Downloadability*
Is there a link to download the whole database?
YES or NO
If YES, include the URL

*Batch*
Is it possible to access the data through a batch feature?
YES or NO

*Query*
Is it possible to access the data through a query-based system?
YES or NO

*Registration*
Finally, is registration compulsory before downloading or accessing data in the ways described above?
YES or NO

Legal accessibility

*Terms of use*
Does the database have a policy?
YES or NO
If YES, include the URL and assess whether the policy authorize reuse, redistribution, integration

Are there any restrictions on the right to reformatting and redistributing?
If NO
If YES, which restrictions?

Fields to describe restrictions are
*Attribution Contractual Requirements*
*Non Derivative Use*
*Non Commercial Use*
*Share Alike*
*Others* (to be described).

---

Fig. 1. Set of questions to process databases

The questions in Figure 1 allow the processing of databases. A subsequent database has been developed with information describing databases technical and legal accessibility.

Five answers can be provided to these questions and constitute a taxonomy to assess technical and legal openness, as presented in Figure 2 below.

---

1. DOWNLOADABILITY
The website provides a file transfer protocol or a link to download the whole dataset without registration.
The ability to download the whole dataset without registration constitute the double requirement to be considered as technically accessible.

2. TECHNICAL RESTRICTION: the database can be accessed only through registration, batch or query-based system.
Technical accessibility is not achieved.

3. PUBLIC DOMAIN POLICY: the website provides simple and clear terms of use informing users that the data are in the public domain.
Data are thus free to integrate. Legal accessibility is achieved.

4. NO POLICY: the website does not provide terms of use.
Legal accessibility is not achieved.

5. LEGAL RESTRICTIONS: the terms of use impose contractual restrictions, such as heavy contractual requirements for attribution, limitation to non-commercial usages, prohibition to modify data, or other constraints on their redistribution or modification.
Legal accessibility is not achieved. The data are not free to integrate.

---

Figure 2. Databases qualification

## 2.2. Databases analysis according to the taxonomy

Samples of the Nucleic Acids Research (NAR) Molecular Biology Database Collection [MBDC](#) and Life Science Resource Name ([LSRN](#)) Schemas databases have been analyzed to define the taxonomy. Then one third of the LSRN databases (60 databases) have been systematically analyzed. A subsequent database has been created, gathering for each of these databases:
- The name and URL of the database,
- URL of the download page and URL of the terms of use,
- Extracts of the terms of use for further review and comments,
- Values for technical accessibility and legal accessibility features as described in Figure 1.

**Technical openness**

Four values have been identified to assess technical accessibility: Downloadability, Batch features, Query-based system and Registration.

The only combination qualifying the database as technically open is the ability to download without registration. Indeed, registration before access and the possibility to perform only batch or query-based searches prevents automated data mining. However, it can be useful to have access to several systems to retrieve and analyze data. Therefore, the database indicates whether it is possible to retrieve data also through batch and query in addition to download.

Another technical restriction that hasn't been analyzed is the presence of standardized annotations or comments allowing users to understand data collected by others. This feature has been disregarded because of a lack of expertise to assess the relevance and quality of annotation for external reuse. A user rating process to evaluate the freshness of the data will be part of the second stage of the project.

**Legal openness**

Values have been used to define legal accessibility are the following: Policy Available, Public Domain Policy, Attribution Contractual Requirements, Non Derivative Use, Non Commercial Use, Share Alike and Others to be described.

In order to be open, a database must have a policy, and this policy must equal to the Public Domain.
The absence of any terms of use or policy on the database website could imply that, in the absence of any expressed restrictions, data are free. But rights unknown to the user might be applicable by default. Indeed, the Protocol states that « any implementation MUST affirmatively declare that contractual constraints do not apply to the database. » Policies should be clear and have only one possible legal interpretation. The absence of a clear and understandable policy is equivalent to the absence of a policy because it leads to legal uncertainty.

Any restriction to the redistribution and the modification of data will prevent the qualification of the policy as Public Domain policy. Legal restrictions to redistribute and modify of data can be diverse. Four values have been identified, corresponding to Creative Commons licenses options: Attribution Contractual Requirements, Non Derivative Use, Non Commercial Use, Share Alike. However, the definition for these legal restrictions in the context of this research is broader than the Creative Commons definitions.
The Attribution requirement may constitute a restriction on the reuse of data. Instead of strong contractually binding requirements on how data should be attributed, a request of acknowledgment according to scientific norms should be sufficient. According to the Protocol, "any implementation SHOULD define a non-legally binding set of citation norms in clear, lay-readable language". Furthermore, "Community standards for sharing publication-related data and materials should flow from the general principle that the publication of scientific information is intended to move science forward. More specifically, the act of publishing is a quid pro quo in which authors receive credit and acknowledgment in exchange for disclosure of their scientific findings. An author's obligation is not only to release data and materials to enable others to verify or replicate published findings (as journals already implicitly or explicitly require) but also to provide them in a form on which other scientists can build with further research."[3]
The Non Commercial and Non Derivative requirements prevent many types of data use. They are defined as restrictions based on the commercial nature of the user or of the usage, and as restrictions on the distribution of modified versions of the database.
The Share Alike requirement is present in the original taxonomy: this option requests modifications to be offered under the same open terms and should gather all copyleft policies.

---

[3] Board on Life Sciences (BLS), Sharing Publication-Related Data and Materials: Responsibilities of Authorship in the Life Sciences (2003). http://books.nap.edu/books/0309088593/html/R1.html

No policy in the analyzed sample contains this requirement. A field should be foreseen for other possible restrictions that may affect terms of use but haven't been identified in the initial subsets. For instance, an embargo on publishing before the data producer, the existence of patents and the absence of warranties against third-party rights are legal restrictions which have not been taken into account in this first analysis. They deserve a subsequent analysis within the scope of databases legal context and issues preventing the freedom to integrate.

In many cases, the database is offered with no restrictions placed by the database curator, but without warranties on the legal status of the data submitted by contributors. Data may contain elements protected by copyright or any applicable right. The database curator did not clear the rights, or did not request from the contributors a rights waiver or no rights assertion before data upload, or does not want to be held liable in case the previously described processes would present a failure. This warranties disclaimer can be seen as a hurdle to the usage of these data. Both uncertainty for the end-user and absence of responsibility for the curator could be avoided by proposing to contributors a data sharing agreement prior to submission, seamlessly integrated in the upload process. Although this procedure might disincentivize some contributors, the burden of checking the legal status of data and avoiding possible claims by third parties should not rely on the data user, forcing her to hire a lawyer. Besides, these disclaimers do not identify which data are free and which parts of the database might be copyrighted or covered by other rights.

## 2.3 Results

Databases which can be considered legally and technically open, and compliant with the Science Commons Open Data Protocol are those that are downloadable without prior registration, under a public domain simple policy. Databases available only through batch or query interfaces are not considered technically open, but those offering these features in addition to downloadability will be compliant.

Besides databases created by the National Center for Biotechnology Information ([NCBI](#)) and the European Bioinformatics Institute ([EBI](#)), only a couple of databases among the 60 first schemas of the LCRN registry analyzed sample can be considered as both technically and legally open, without restrictions. These are almost exclusively governmental databases.

NCBI is part of the US National Institutes of Health and the EBI is part of European Molecular Biology Laboratory (EMBL), funded by its 20 member states, as well as the European Commission, Wellcome Trust, US National Institutes of Health, UK Research Councils, our industry partners and the UK Department of Trade and Industry. Not only these institutions are pioneers in bioinformatics and genetics, but also in setting Open Access policies. Indeed, the [NIH Public Access Policy](#) requires "scientists to submit journal final peer-reviewed manuscripts that arise from NIH funds to the digital archive [PubMed Central](#) (…) to help advance science and improve human health". The European Research Council Scientific Council Guidelines for Open Access also "requires that all peer-reviewed publications from ERC-funded research projects be deposited on publication into an appropriate research repository", but only "considers essential that primary data - which in the life sciences for example could comprise data such as nucleotide/protein sequences, macromolecular atomic coordinates and anonymized epidemiological data - are deposited to the relevant databases."

It seems that the next challenge for Open Access will be to analyze barriers and incentives to the deposit of scientific data in open repositories.

**3. A checklist to assess databases openness**

We propose a checklist to assist data curators in opening their data, and to make sure that the database design and terms of use will allow others to access, reuse and build upon their data. All answers should be positive.

---

*A. Check your database technical accessibility*

A.1. Do you provide a link to download the whole database?
A.2. Is the dataset available in at least one standard format?
A.3. Do you provide comments and annotations fields allowing users to understand the data?

*B. Check your database legal accessibility*

B.1. Do you provide a policy expressing terms of use of your database?
B.2. Is the policy clearly indicated on your website?
B.3. Are the terms short and easy to understand by non-lawyers?
B.4. Does the policy authorize redistribution, reuse and modification without restrictions or contractual requirements on the user or the usage?
B.5. Is the attribution requirement at most as strong as the acknowledgment norms of your scientific community?

---

Figure 3. Database openness checklist

**4. Conclusion**

This article presents technical and legal restrictions typically applied to life science databases based on the analysis of two samples of terms of use. Most of these terms of use impose legal restrictions on the redistribution of data. The absence of policy, or of clear and understandable terms, also prevents reuse by the lack of legal certainty. Besides legal restrictions, many databases impose technical constraints on their users. The impossibility or the difficulty of downloading and reformatting the dataset does not fulfill technical accessibility requirements. Databases available only through batch or query interfaces are not considered technically open, but those offering these features in addition to downloadability will be compliant. Databases that can be considered legally and technically open, and compliant with the Science Commons Open Data Protocol are those that are downloadable without prior registration, under a public domain simple policy.

**5. Future work**

A [user interface](#) has been built to host the dataset assessing databases technical and legal accessibility. A search function allows the classification of databases according to their accessibility status and the identification of those that are compliant with the protocol.
The dataset analyzing databases' terms of use with regard to the taxonomy can now be opened to other contributors in order to gather more results and assess the level of restrictions of more databases in the field. Statistical analysis may be performed on those data to assess the degree of openness on a given databases sregistry. It would also be interesting to know whether the taxonomy can be extended to other disciplines.

This dataset of legal and technical controls can also serve as a repository for commenting on databases policies. The taxonomy allows search according to technical and legal restrictions, and this can provide researchers clear information on what they can and cannot do.
It is also possible to implement these restrictions into metadata expressed with the ccREL standard, and turn the taxonomy into a formalized ontology of database freedoms. It is also possible to develop the website into a repository similar to OAKlist or the Sherpa/Romeo archiving scientific publishers copyright self-archiving policies. Classifying databases according to their access and reuse policy status may help scientists and open access scholars understand and appreciate which databases are really open without having to read all the policies, and maybe databases curators to further open their data.

## 6. Acknowledgments

This work has been developed within the Science Commons Data project during my 2007-2008 fellowship at Science Commons. I thank John Wilbanks for inspiration and support, and Shirley Fung for designing and developing the user interface hosting the database. I am also grateful to Carolina Rossini for analyzing some databases with me. I was expecting that two opinions on the terms of use would serve as a review of the categorization according to the taxonomy legal accessibility values. However, this experience proved to be more difficult than expected, demonstrating that if terms of use are not uniformly understood and categorized by two persons with legal expertise, they cannot be considered clear policies. Each one of them should all be thanked for reviewing the taxonomy and the paper at different stages of the project. Thinh Nguyen, Jonathan Rees and Victoria Stodden should also be thanked for their comments. Finally, this work would not have been possible without the database curated by Life Science Resource Name (LSRN).

## 7. References

Life Science Resource Name (LSRN), http://lsrn.org/lsrn/registry.html

Nucleic Acids Research (NAR) Molecular Biology Database Collection (MBDC)
http://nar.oxfordjournals.org/cgi/content/full/gkm1037/DC1/1

Sherpa/Romeo repository, http://www.sherpa.ac.uk/romeo.php

User interface of the project, http://shirleyfung.com/mbdb/

9